\documentclass[conference,a4paper]{APSIPA2019}
\usepackage{multirow}
\usepackage[dvips]{graphicx}
\usepackage{amsmath}
\usepackage[psamsfonts]{amssymb}
\usepackage{amsxtra}
\usepackage{threeparttable}

\begin{document}

\title{AP20-OLR Challenge: Three Tasks and Their Baselines}

\author{%
\authorblockN{%
Zheng Li\authorrefmark{2},
Miao Zhao\authorrefmark{3},
Qingyang Hong\authorrefmark{3}$^*$,
Lin Li\authorrefmark{2}$^*$,
Zhiyuan Tang\authorrefmark{4},
Dong Wang\authorrefmark{4}\authorrefmark{5},
\\
Liming Song\authorrefmark{8} and
Cheng Yang\authorrefmark{8}
}
\authorblockA{%
\authorrefmark{2}
School of Electronic Science and Engineering, Xiamen University\\
}
\authorblockA{%
\authorrefmark{3}
School of Informatics, Xiamen University\\
}
\authorblockA{%
\authorrefmark{4}
Center for Speech and Language Technologies, Tsinghua University\\
}
\authorblockA{%
\authorrefmark{5}
Beijing National Research Center for Information Science and Technology\\
\authorrefmark{8}
Speechocean\\
Corresponding email:  qyhong@xmu.edu.cn, lilin@xmu.edu.cn
}
}

\maketitle
\thispagestyle{empty}

\begin{abstract}
This paper introduces the fifth oriental language recognition (OLR) challenge AP20-OLR, which intends to improve the performance of language recognition systems, along with APSIPA Annual Summit and Conference (APSIPA ASC). The data profile, three tasks, the corresponding baselines, and the evaluation principles are introduced in this paper.

The AP20-OLR challenge includes more languages, dialects and real-life data provided by Speechocean and the NSFC M2ASR project, and all the data is free for participants. The challenge this year still focuses on practical and challenging problems, with three tasks: (1) cross-channel LID, (2) dialect identification and (3) noisy LID. Based on Kaldi and Pytorch, recipes for i-vector and x-vector systems are also conducted as baselines for the three tasks. These recipes will be online-published, and available for participants to configure LID systems. The baseline results on the three tasks demonstrate that those tasks in this challenge are worth paying more efforts to achieve better performance.

  \end{abstract}


  \section{Introduction}

  The language identification (LID) refers to identify the language categories from utterances, and it is usually presented at the front end of speech processing systems, such as the automatic speech recognition (ASR), meaning that the LID technology plays a great role in the applications of multilingual interaction.
  However, there are still difficult issues, decaying the performance of LID systems, such as  the cross-channel issue, the lack of training resources condition and the noisy environment.

  The oriental language families, as a part of many language families around the world, often include Austroasiatic languages (e.g.,Vietnamese, Cambodia)~\cite{sidwell201114},
  Tai-Kadai languages (e.g., Thai, Lao), Hmong-Mien languages (e.g., some dialects in south China), Sino-Tibetan languages (e.g., Chinese Mandarin), Altaic languages (e.g., Korea, Japanese) and Indo-European languages (e.g., Russian)~\cite{ramsey1987languages,shibatani1990languages,comrie1996russian}.

  Dialect, often referring to a variety of a specific language,
  is also a typical linguistic phenomenon.
  Different dialects may be considered as different kinds of languages for speech processing.
  As an oriental country, China has 56 ethnic groups, and each ethnic group has its own unique dialect(s).
  Some of these Chinese dialects may share some part of written system with Mandarin Chinese,
  but the totally different pronunciation results in more complicated multilingual phenomena.
  In the community, recently there was an Arabic dialect ID task in MG5-5 challenge\cite{MG5}.


  The oriental language recognition (OLR) challenge is organized annually,
  aiming at improving the research on multilingual phenomena and
  advancing the development of language recognition technologies.
  The challenge has been conducted four times since 2016,
  namely AP16-OLR~\cite{wang2016ap16}, AP17-OLR~\cite{tang2017ap17}, AP18-OLR~\cite{tang2018ap18} and AP19-OLR~\cite{tang2019ap19},
  each attracting dozens of teams around the world.

  AP19-OLR involved more than $10$ languages and focused on three challenging tasks:
  (1) short-utterance ($1$ second) LID, which was inherited from AP18-OLR;
  (2) cross-channel LID; (3) zero-resource LID.
  In the first task, the system submitted by the Innovem-Tech team achieved the best performance
  ($C_{avg}$ with $0.0212$, and EER with $2.47\%$).
  In the second task, the system submitted by the Samsung SSLab team achieved the best $C_{avg}$  performance
  with $0.2008$, and EER with $20.24\%$. And in the third task, the system submitted by the XMUSPEECH team achieved the best $C_{avg}$ performance with $0.0113$, and EER with $1.13\%$.
  From these results, one can see that for the cross-channel condition, the task remains
  challenging.  More details about the past four challenges can be found
  on the challenge website.\footnote{http://olr.cslt.org}

  Based on the experience of the last four challenges and the calling from industrial application,
  we propose the fifth OLR challenge.
  This new challenge, denoted by AP20-OLR, will be hosted by APSIPA ASC 2020.
  It involves more languages/dialects and focuses on more practical and challenging tasks:
  (1) cross-channel LID, as in the last challenge,
  (2) dialect identification,  where three dialect resources are provided for training,
  but other three languages are also included in the test set, to compose the open-set dialect identification, and
  (3) noisy LID, which reveals another real-life demand of speech technology to deal with the low SNR condition.

In the rest of the paper, we will present the data profile and the evaluation plan of the AP20-OLR challenge. To assist participants to build their own submissions, two types of baseline systems are provided, based on Kaldi and Pytorch respectively.

  \begin{table*}[htb]
  \begin{center}
  \caption{AP16-OL7 and AP17-OL3 Data Profile}
  \label{tab:ol10}
  \begin{tabular}{|l|l|c|c|c|c|c|c|c|}
   \hline
  \multicolumn{3}{|c|}{\textbf{AP16-OL7}} & \multicolumn{3}{c|}{AP16-OL7-train/dev}  & \multicolumn{3}{c|}{AP16-OL7-test}\\
  \hline
  Code & Description & Channel & No. of Speakers & Utt./Spk. & Total Utt. & No. of Speakers & Utt./Spk. & Total Utt. \\
  \hline
  ct-cn & Cantonese in China Mainland and Hongkong & Mobile & 24 & 320 & 7559 & 6 & 300 & 1800 \\
  \hline
  zh-cn & Mandarin in China & Mobile & 24 & 300 & 7198        & 6 & 300 & 1800 \\
  \hline
  id-id & Indonesian in Indonesia &  Mobile & 24 & 320 & 7671 & 6 & 300 & 1800 \\
  \hline
  ja-jp & Japanese in Japan & Mobile & 24 & 320 & 7662        & 6 & 300 & 1800 \\
  \hline
  ru-ru & Russian in Russia & Mobile & 24 & 300 & 7190        & 6 & 300 & 1800 \\
  \hline
  ko-kr & Korean in Korea & Mobile & 24 & 300 & 7196          & 6 & 300 & 1800 \\
  \hline
  vi-vn & Vietnamese in Vietnam & Mobile & 24 & 300 & 7200    & 6 & 300 & 1800 \\
   \hline
  \hline
   \multicolumn{3}{|c|}{\textbf{AP17-OL3}} & \multicolumn{3}{c|}{AP17-OL3-train/dev}  & \multicolumn{3}{c|}{AP17-OL3-test}\\
  \hline
  Code & Description & Channel & No. of Speakers & Utt./Spk. & Total Utt. & No. of Speakers & Utt./Spk. & Total Utt. \\
  \hline
  ka-cn & Kazakh in China & Mobile & 86 & 50  & 4200 &      86 &  20  & 1800 \\
  \hline
  ti-cn & Tibetan in China & Mobile & 34 & 330   & 11100 &    34 & 50  & 1800 \\
  \hline
  uy-id & Uyghur in China &  Mobile & 353 & 20   & 5800 &    353 & 5  & 1800 \\
   \hline
  \end{tabular}
  \begin{tablenotes}
  \item[a] Male and Female speakers are balanced.
  \item[b] The number of total utterances might be slightly smaller than expected, due to the quality check.
  \end{tablenotes}
  \end{center}
  \end{table*}

  \section{Database profile}

  Participants of AP20-OLR can request the following datasets for system construction.
  All these data can be used to train their submission systems as follows.

  \begin{itemize}
  \item AP16-OL7: The standard database for AP16-OLR, including AP16-OL7-train, AP16-OL7-dev and AP16-OL7-test.
  \item AP17-OL3: A dataset provided by the M2ASR project, involving three new languages. It contains AP17-OL3-train and AP17-OL3-dev.
  \item AP17-OLR-test: The standard test set for AP17-OLR. It contains AP17-OL7-test and AP17-OL3-test.
  \item AP18-OLR-test: The standard test set for AP18-OLR. It contains AP18-OL7-test and AP18-OL3-test.
  \item AP19-OLR-dev: The development set for AP19-OLR. It contains AP19-OLR-dev-task2 and AP19-OLR-dev-task3.
  \item AP19-OLR-test: The standard test set for AP19-OLR. It contains AP19-OL7-test and AP19-OL3-test.
  \item AP20-OLR-dialect: The newly provided training set, including three kinds of Chinese dialects.
  \item THCHS30:  The THCHS30 database (plus the accompanied resources) published by CSLT, Tsinghua University~\cite{wang2015thchs}.
  \end{itemize}


  Besides the speech signals, the AP16-OL7 and AP17-OL3 databases also provide lexicons of all the 10 languages, as well
  as the transcriptions of all the training utterances. These resources allow training acoustic-based or phonetic-based
  language recognition systems. Training phone-based speech recognition systems is also possible, though
  large vocabulary recognition systems are not well supported, due to the lack of large-scale language models.

  A test dataset AP20-OLR-test will be provided at the date of result submission,
  which includes three parts corresponding to the three LID tasks.


  \subsection{AP16-OL7}

  The AP16-OL7 database was originally created by Speechocean, targeting for various speech processing tasks.
  It was provided as the standard training and test data in AP16-OLR.
  The entire database involves 7 datasets, each in a particular language. The seven languages are:
  Mandarin, Cantonese, Indonesian, Japanese, Russian, Korean and Vietnamese.
  The data volume for each language is about $10$ hours of speech signals recorded in
  reading style. The signals were
  recorded by mobile phones, with a sampling rate of $16$ kHz  and a sample size of $16$ bits.

  For Mandarin, Cantonese, Vietnamese and Indonesia, the recording was conducted in a quiet environment.
  As for Russian, Korean and Japanese, there are $2$ recording sessions for each speaker: the first session
  was recorded in a quiet environment and the second was recorded in a noisy environment.
  The basic information of the AP16-OL7 database is presented in Table~\ref{tab:ol10},
  and the details of the database can be found in the challenge website or the
  description paper~\cite{wang2016ap16}.

  \subsection{AP17-OL7-test}

  The AP17-OL7 database is a dataset provided by SpeechOcean. This dataset contains 7 languages as in AP16-OL7,
  each containing $1800$ utterances. The recording conditions are the same as AP16-OL7. This database is used as
  part of the test set for the AP17-OLR challenge.

  \subsection{AP17-OL3}

  The AP17-OL3 database contains 3 languages: Kazakh, Tibetan and Uyghur, all are minority languages in China.
  This database is part of the Multilingual Minorlingual Automatic Speech Recognition (M2ASR) project, which is
  supported by the National Natural Science Foundation of China (NSFC). The project is a three-party collaboration, including Tsinghua University,
  the Northwest National University, and Xinjiang University~\cite{wangm2asr}. The aim of this project is to construct speech recognition systems for five minor languages in China (Kazakh, Kirgiz, Mongolia, Tibetan and Uyghur). However, our ambition is beyond that scope: we hope
  to construct a full set of linguistic and speech resources and tools for the five languages, and make them open and free for
  research purposes. We call this the M2ASR Free Data Program. All the data resources, including the tools published in this paper, are released on the web site of the project.\footnote{http://m2asr.cslt.org}

  The sentences of each language in AP17-OL3 are randomly selected from the original M2ASR corpus.
  The data volume for each language in AP17-OL3 is about $10$ hours of speech signals
  recorded in reading style.
  The signals were recorded by mobile phones,
  with a sampling rate of $16$ kHz and a sample size of $16$ bits.
  We selected $1800$ utterances for each language as the development set (AP17-OL3-dev), and the rest is used as the
  training set (AP17-OL3-train). The test set of each language involves $1800$ utterances, and is provided separately
  and denoted by AP17-OL3-test.
  Compared to AP16-OL7, AP17-OL3 contains much more variations in terms of recording conditions and
  the number of speakers, which may inevitably  increase the difficulty of the challenge task.
  The information of the AP17-OL3 database is summarized in Table~\ref{tab:ol10}.

  \subsection{AP18-OLR-test}
  The AP18-OLR-test database is the standard test set for AP18-OLR,
  which contains AP18-OL7-test and AP18-OL3-test.
  Like the AP17-OL7-test database,
  AP18-OL7-test contains the same target $7$ languages, each containing $1800$ utterances,
  while AP18-OL7-test also contains utterances
  from several interference languages.
  The recording conditions are the same as AP17-OL7-test.
  Like the AP17-OL3-test database,
  AP18-OL3-test contains the same $3$ languages, each containing $1800$ utterances.
  The recording conditions are also the same as AP17-OL7-test.

  \subsection{AP19-OLR-test}
  The AP19-OLR-test database is the standard test set for AP19-OLR,
  which includes 3 parts responding to the 3 LID tasks respectively, precisely
  AP19-OLR-short, AP19-OLR-channel and AP19-OLR-zero.

  \subsection{AP20-OLR-dialect}
  AP20-OLR-dialect is the training set provided by SpeechOcean. It includes three kinds of Chinese dialects, namely Hokkien, Sichuanese and Shanghainese. The utterances of each language are about 8000. The signals were
  recorded by mobile phones, with a sampling rate of $16$ kHz  and a sample size of $16$ bits.

  \subsection{AP20-OLR-test}
  The AP20-OLR-test database is the standard test set for AP20-OLR, which includes 3 parts responding to the 3 LID tasks respectively, precisely AP20-OLR-channel-test, AP20-OLR-dialect-test and AP20-OLR-noisy-test.

    \begin{itemize}
  \item AP20-OLR-channel-test: This subset is designed for the cross-channel LID task, which contains six of the ten target languages, but was recorded with different recording equipments and environment. The six languages are Cantonese,  Indonesian, Japanese, Russian, Korean and Vietnamese. Each language has about 1800 utterances.

  \item AP20-OLR-dialect-test: This subset is designed for the dialect identification task, including three dialects which are Hokkien, Sichuanese and Shanghainese. Considering the real-life situation, other three kinds of nontarget languages, which are Mandarin, Malay and Thai, are included in this subset to compose the open-set dialect identification, and there may be some cross channel utterances as well. Each dialect/language has about 1800 utterances.

  \item AP20-OLR-noisy-test: This subset is designed for the noisy LID task, which contains five of the ten target languages, but was recorded under noisy environment (low SNR). The five languages are Cantonese,   Japanese, Russian, Korean and Mandarin. Each language has about 1800 utterances.
  \end{itemize}

  \section{AP20-OLR challenge}

  Following the definition of NIST LRE15~\cite{lre15}, the task of the LID challenge is defined
  as follows: Given  a  segment  of  speech  and  a  language  hypothesis (i.e.,  a  target
  language  of  interest  to  be  detected),  the  task  is  to decide  whether  that
  target  language  was  in  fact  spoken  in  the given segment (yes or no), based on an
  automated analysis of the data contained in the segment.



  The AP20-OLR challenge includes three tasks as follows:

  \begin{itemize}
  \item Task 1: cross-channel LID  is a close-set identification task, which means the language of each utterance is among the known traditional $6$ target languages, but utterances were recorded with different channels.

  \item Task 2: dialect identification is a open-set identification task, in which three nontarget languages are added to the test set with the three target dialects.

  \item Task 3: noisy LID, where noisy test data of the $5$ target languages will be provided.
  \end{itemize}

  \subsection{System input/output}

  The input to the LID system is a set of speech segments in unknown languages.
  For task 1 and task 3, those speech segments are within
  the $6$ or $5$ known target languages.
  For task 2, the three target dialects of the speech segments are the same as three dialects in the AP20-OLR-dialect.
  The task of the LID system is to determine
  the confidence that a language is contained in a speech segment. More specifically,
  for each speech segment, the LID system outputs a score vector $<\ell_1, \ell_2, ..., \ell_{10}>$,
  where $\ell_i$ represents the confidence that language $i$ is spoken in the speech segment.
  The scores should be comparable across languages and segments.
  This is consistent with
  the principles of LRE15, but differs from that of LRE09~\cite{lre09} where an explicit decision
  is required for each trial.

  In summary, the output of an OLR submission will be a text file, where each line contains
  a speech segment plus a score vector for this segment, e.g.,

  \vspace{0.5cm}
  \begin{tabular}{ccccccccc}
          & lang$_1$   & lang$_2$   & ... & lang$_9$  & lang$_{10}$\\
  seg$_1$ & 0.5  & -0.2 &  ...& -0.3 & 0.1    \\
  seg$_2$ & -0.1 & -0.3 &  ...& 0.5 & 0.3    \\
  ...   &      &     &  ... &      &
  \end{tabular}

  \subsection{Training condition}

    \begin{itemize}
  \item The use of additional training materials is forbidden, including the use of non-speech data for data augmentation purposes. The only resources that are allowed to use are:
        AP16-OL7, AP17-OL3, AP17-OLR-test, AP18-OLR-test, AP19-OLR-test, AP19-OLR-dev, AP20-OLR-dialect and THCHS30.
          \end{itemize}

  \subsection{Test condition}


  \begin{itemize}
  \item All the trials should be processed. Scores of lost trials will be interpreted as -$\inf$.
  \item The speech segments in each task
        should be processed independently, and each test segment in a group should be processed
        independently too. Knowledge from other test segments is not allowed to use (e.g.,
        score distribution of all the test segments).
  \item Information of speakers is not allowed to use.
  \item Listening to any speech segments is not allowed.
  \end{itemize}

  \subsection{Evaluation metrics}

  As in LRE15, the AP20-OLR challenge chooses $C_{avg}$ as the principle evaluation metric.
  First define the pair-wise loss that composes the missing and
  false alarm probabilities for a particular target/non-target language pair:

  \[
  C(L_t, L_n)=P_{Target} P_{Miss}(L_t) + (1-P_{Target}) P_{FA}(L_t, L_n)
 \]

  \noindent where $L_t$ and $L_n$ are the target and non-target languages, respectively; $P_{Miss}$ and
  $P_{FA}$ are the missing and false alarm probabilities, respectively. $P_{Target}$ is the prior
  probability for the target language, which is set to $0.5$ in the evaluation. Then the principle metric
  $C_{avg}$ is defined as the average of the above pair-wise performance:


  \[
   C_{avg} = \frac{1}{N} \sum_{L_t} \left\{
  \begin{aligned}
    & \ P_{Target} \cdot P_{Miss}(L_t) \\
    &  + \sum_{L_n}\ P_{Non-Target} \cdot P_{FA}(L_t, L_n)\
  \end{aligned}
  \right\}
  \]

  \noindent where $N$ is the number of languages, and $P_{Non-Target}$ = $(1-P_{Target}) / (N -1 )$.
  For the open-set testing condition, all of interfering languages will be seen as one unknown language in the computation of $C_{avg}$. 
  We have provided the evaluation scripts for system development.

  \section{Baseline systems}

Two kinds of baseline LID systems were constructed in this challenge based on Kaldi \cite{povey2011kaldi} and Pytorch \cite{Pytorch}: the i-vector model baseline and the extended TDNN x-vector model baselines, respectively. The feature extracting and back-ends were all conducted with Kaldi. To provide more options, we built the i-vector and x-vector models with Kaldi, and conducted an x-vector model with Pytorch as well. The Kaldi and Pytorch recipes of these baselines can be downloaded from the challenge web site.\footnote{http://cslt.riit.tsinghua.edu.cn/mediawiki/index.php/OLR\_Challenge\_2020}

  We trained the baseline systems with a combined dataset including AP16-OL7, AP17-OL3 and AP17-OLR-test, and the target number of the system refers to the number of all languages, i.e. $10$. Before training, we adopted the data augmentation, including speed and volume perturbation, to increase the amount and diversity of the training data. For speed perturbation, we applied a speed factor of 0.9 or 1.1 to slow down or speed up the original recording. And for volume perturbation, random volume factor was applied. Finally, two augmented copies of the original recording were added to the original data set to obtain a 3-fold combined training set.

  The acoustic features involved 20-dimensional Mel frequency cepstral coefficients (MFCCs) with the 3-dimensional pitch, and the energy VAD was used to filter out nonspeech frames.

  The back-end was the same for all three tasks when the embeddings was extracted from the model. Linear discriminative analysis (LDA) trained on the enrollment set was employed to promote language-related information. The dimensionality of the LDA projection space was set to 100. After the LDA projection and centering, the logistic regression (LR) trained on the enrollment set was used to compute the score of a trial on a particular language.

  \subsection{i-vector system}

The i-vector baseline system was constructed based on the i-vector model [15], [16]. The acoustic features were augmented by their first and second order derivatives, resulting in 69-dimensional feature vectors. The UBM involved 2,048 Gaussian components and the dimensionality of the i-vectors was 600.

  \subsection{x-vector system}

  We used the x-vector system with extended TDNN as the x-vector baseline system \cite{snyder2018x,snyder2018spoken,villalba2016etdnn}.
  Compared to the traditional x-vector, the extended TDNN x-vector structure used a slightly wider temporal context in the TDNN layers and interleave dense layers between TDNN layers, which leaded to a deeper x-vector model. This deep structure was trained to classify the $N$ languages in the training data with the cross entropy (CE) loss. After training, embeddings called `x-vector' were extracted from the affine component of the penultimate layer. Two implementations of this model were conducted on Kaldi and Pytorch, respectively.

   \subsubsection{Implementation details on Kaldi}

   The chunk size between 60 to 80 was used in the sequential sampling when prepared the training examples. The model was optimized with SGD optimizer, with a mini-batch size of 128. The Kaldi's parallel training and sub-models fusion strategy was used.

  \begin{table}[b]
\begin{center}
\setlength{\belowcaptionskip}{-0pt}
\caption{C$_{avg}$ and EER results on the referenced development sets}
\label{tab:1}
\begin{tabular}{cccccc}
\cline{1-5}
\multicolumn{1}{|c|}{Task}                                                             & \multicolumn{2}{c|}{Cross-channel LID}                 & \multicolumn{2}{c|}{Dialect Identification}                     &  \\ \cline{1-5}
\multicolumn{1}{|c|}{Enrollment Set}                                                   & \multicolumn{2}{c|}{AP20-ref-dev-task1}                 & \multicolumn{2}{c|}{AP20-OLR-dialect}                           &  \\ \cline{1-5}
\multicolumn{1}{|c|}{Test Set}                                                         & \multicolumn{2}{c|}{AP19-OLR-channel}                          & \multicolumn{2}{c|}{AP19-OLR-dev\&eval-task3-test}                     &  \\ \cline{1-5}
\multicolumn{1}{|c|}{}                                                                 & \multicolumn{1}{c|}{\textit{Cavg}} & \multicolumn{1}{c|}{EER\%} & \multicolumn{1}{c|}{\textit{Cavg}} & \multicolumn{1}{c|}{EER\%} &  \\ \cline{1-5}
\multicolumn{1}{|c|}{\begin{tabular}[c]{@{}c@{}}{[}Kaldi{]}\\ i-vector\end{tabular}}   & \multicolumn{1}{c|}{0.2965}        & \multicolumn{1}{c|}{29.12} & \multicolumn{1}{c|}{0.0703}        & \multicolumn{1}{c|}{9.33}  &  \\ \cline{1-5}
\multicolumn{1}{|c|}{\begin{tabular}[c]{@{}c@{}}{[}Kaldi{]}\\ x-vector\end{tabular}} & \multicolumn{1}{c|}{0.3583}        & \multicolumn{1}{c|}{36.37} & \multicolumn{1}{c|}{0.0807}        & \multicolumn{1}{c|}{14.67}  &  \\ \cline{1-5}
\multicolumn{1}{|c|}{\begin{tabular}[c]{@{}c@{}}{[}Pytorch{]}\\ x-vector\end{tabular}}   & \multicolumn{1}{c|}{0.2696}        & \multicolumn{1}{c|}{26.94} & \multicolumn{1}{c|}{0.0849}        & \multicolumn{1}{c|}{12.40}  &  \\ \cline{1-5}
                                                                                       &                                    &                            &                                    &                            &
\end{tabular}
    \end{center}
\end{table}

\begin{table*}[hbt]
\begin{center}
\setlength{\belowcaptionskip}{-30pt}
\caption{C$_{avg}$ and EER results on the ap20 evaluation sets}
\label{tab:2}
\begin{tabular}{ccccccc}
\hline
\multicolumn{1}{|c|}{Task}                                                             & \multicolumn{2}{c|}{Cross-channel LID}                          & \multicolumn{2}{c|}{Dialect Identification}                     & \multicolumn{2}{c|}{Noisy LID}                                  \\ \hline
\multicolumn{1}{|c|}{Enrollment Set}                                                   & \multicolumn{2}{c|}{AP20-ref-enroll-task1}                 & \multicolumn{2}{c|}{AP20-OLR-dialect}                           & \multicolumn{2}{c|}{AP20-ref-enroll-task3}                \\ \hline
\multicolumn{1}{|c|}{Test Set}                                                         & \multicolumn{2}{c|}{AP20-OLR-channel-test}                          & \multicolumn{2}{c|}{ AP20-OLR-dialect-test}                          & \multicolumn{2}{c|}{AP20-OLR-noisy-test}                          \\ \hline
\multicolumn{1}{|c|}{}                                                                 & \multicolumn{1}{c|}{\textit{Cavg}} & \multicolumn{1}{c|}{EER\%} & \multicolumn{1}{c|}{\textit{Cavg}} & \multicolumn{1}{c|}{EER\%} & \multicolumn{1}{c|}{\textit{Cavg}} & \multicolumn{1}{c|}{EER\%} \\ \hline
\multicolumn{1}{|c|}{\begin{tabular}[c]{@{}c@{}}{[}Kaldi{]}\\ i-vector\end{tabular}}   & \multicolumn{1}{c|}{0.1542}        & \multicolumn{1}{c|}{19.40} & \multicolumn{1}{c|}{0.2214}        & \multicolumn{1}{c|}{23.94} & \multicolumn{1}{c|}{0.0967
}        & \multicolumn{1}{c|}{9.77} \\ \hline
\multicolumn{1}{|c|}{\begin{tabular}[c]{@{}c@{}}{[}Kaldi{]}\\ x-vector\end{tabular}} & \multicolumn{1}{c|}{0.2098}        & \multicolumn{1}{c|}{22.49} & \multicolumn{1}{c|}{0.2117}        & \multicolumn{1}{c|}{22.25} & \multicolumn{1}{c|}{0.1079}        & \multicolumn{1}{c|}{11.12}  \\ \hline
\multicolumn{1}{|c|}{\begin{tabular}[c]{@{}c@{}}{[}Pytorch{]}\\ x-vector\end{tabular}}   & \multicolumn{1}{c|}{0.1321}        & \multicolumn{1}{c|}{14.58} & \multicolumn{1}{c|}{0.1752}        & \multicolumn{1}{c|}{19.74} & \multicolumn{1}{c|}{0.0715}         & \multicolumn{1}{c|}{7.14} \\ \hline
                                                                                       &                                    &                            &                                    &                            &                                    &
\end{tabular}
    \end{center}
\end{table*}

   \subsubsection{Implementation details on Pytorch}

   The chunk size was 100 with the language-balanced sampling when prepared the training examples. The language-balanced sampling ensured that the examples in languages were roughly the same, by the repeated sampling of languages with less training frames. The model was optimized with Adam optimizer, with a mini-batch size of 512. The warm restarts was used to control the learning rate and the feature dropout was used to enhance the robustness.

  \subsection{Performance results}

  The primary evaluation metric in AP20-OLR is $C_{avg}$. Besides that, we also present the performance
  in terms of equal error rate (EER). These metrics evaluate
  system performance from different perspectives, offering a whole picture of the capability
  of the tested system.  The performance of baselines is evaluated on the AP20-OLR-test database, but we also provide the
  results on the referenced development sets. For task 1, we choose the cross-channel subset of AP19-OLR-test to be the referenced development set. For task 2, the dialect test subset of AP19-OLR-dev-task3, which contains three target dialects: Hokkien, Sichuanese and Shanghainese, and the test subset of AP19-OLR-eval-task3, which contains three nontarget (interfering) languages: Catalan, Greek and Telugu, are combined as the referenced development set. While no noisy data set was given in the past OLR challenges, we do not present the referenced development set of task 3 in this challenge. The referenced development sets are used to help estimate the system performance when participants reproduce the baseline systems or prepare their own systems, and participants are encouraged to design their own development sets.

  Table~\ref{tab:1} and Table ~\ref{tab:2} show the utterance-level C$_{avg}$ and EER results on the referenced development sets and AP20-OLR-test, respectively.

  \subsubsection{Cross-channel LID}

  The first task identifies cross-channel utterances.
  The enrollment sets are subsets of the 3-fold combined training set mentioned above, in which the utterances of the same languages in test sets are reserved, namely AP20-ref-dev-task1 and AP20-ref-enroll-task1.
  In the referenced development set, the inconsistency of two metrics $C_{avg}$ and EER is observed. The x-vector model on Pytorch achieves the best $C_{avg}$ performance with $0.2696$, but the EER performance with $26.94\%$, which was much worse than i-vector model (EER with $19.40\%$). In the evaluation set, the x-vector model on Pytorch achieves the best $C_{avg}$ performance with $0.1321$, and EER with $14.58\%$. Meanwhile, the trend of baselines' performance is different between the referenced development set and the evaluation set, since the channel conditions might be different between these two test sets.

  \subsubsection{Dialect Identification}
  The second task identifies three target dialects from six languages/dialects. In the referenced development set, it should be noted that the AP19-OLR-dev-task3-test only contain 500 utterances per target dialects, and the two interfering languages in the AP19-OLR-eval-task3-test are European languages, while the languages in the training set are all oriental languages, so the performance of Pytorch's x-vector baseline is relatively unsatisfactory, since it's stronger fitting of the training data, comparing with the Kaldi's model. In the evaluation set, three nontarget languages are Asian languages, and the best performance on  $C_{avg}$  is achieved on the Pytorch's x-vector model, with $0.1752$, and EER with $19.74\%$. The Kaldi's x-vector and i-vector models have similar performance in both the referenced development set and the evaluation set in terms of $C_{avg}$.



  \subsubsection{Noisy LID}
  The evaluation process for task 3 can be seen as identifying languages from 5 target languages under noisy testing condition. No referenced development is given for this task. The referenced enrollment set is subsets of the 3-fold combined training set mentioned above, in which the utterances of the same languages as in test sets were reserved, namely AP20-ref-enroll-task3. In the evaluation set,
  the x-vector model on Pytorch achieves the best performance of $C_{avg}$ and EER, with $0.0715$, and $7.14\%$, respectively. The performance of Kaldi's x-vector and i-vector models is close in the evaluation set.


  \section{Conclusions}

  In this paper, we present the data profile, three task definitions and their baselines of the AP20-OLR challenge. In this challenge, besides the presented data sets in past challenges, new dialect training/developing data sets are provided for participants, and more languages are included in the test sets. The AP20-OLR challenge are with three tasks: (1) cross-channel LID, (2) dialect identification and (3) noisy LID. The i-vector and x-vector frameworks, deploying with Kaldi and Pytorch, are conducted as baseline
  systems to assist participants to construct a reasonable starting system. Given the results on baseline systems, these three tasks defined by AP20-OLR are rather challenging and are worthy of careful study.
  All the data resources are free for the participants, and the recipes of the baseline systems can
  be freely downloaded.


  \section*{Acknowledgment}

  This work was supported by the National Natural Science
  Foundation of China No.61876160 and No.61633013.

  We would like to thank Ming Li at Duke Kunshan University, Xiaolei Zhang at  Northwestern
  Polytechnical University for their help in organizing this AP20-OLR challenge.

  \bibliographystyle{IEEEtran}
  \bibliography{olr}

  \end{document}